\documentclass[11pt,onecolumn]{IEEEtran}

\usepackage{amsmath,amssymb,amsfonts}
\usepackage{graphicx}
\usepackage{textcomp}

\usepackage{subcaption}

\usepackage{epsfig, epsf, array, latexsym, graphics, multirow, color}
\usepackage{booktabs}
\usepackage{multirow}
\usepackage{multicol}
\usepackage{colortbl}
\usepackage{graphicx}
\usepackage{amsmath}
\usepackage{algorithm}
\usepackage{algpseudocode}

\newtheorem{definition}{Definition}
\newtheorem{proposition}{Proposition}
\newtheorem{theorem}{Theorem}

\newcommand{\ls}[1]
   {\dimen0=\fontdimen6\the\font
    \lineskip=#1\dimen0
    \advance\lineskip.5\fontdimen5\the\font
    \advance\lineskip-\dimen0
    \lineskiplimit=.9\lineskip
    \baselineskip=\lineskip
    \advance\baselineskip\dimen0
    \normallineskip\lineskip
    \normallineskiplimit\lineskiplimit
    \normalbaselineskip\baselineskip
    \ignorespaces
   }

\newcommand {\bearn}{\begin{eqnarray*}}
\newcommand {\eearn}{\end{eqnarray*}}
\newcommand {\barr}{\begin{array}}
\newcommand {\earr}{\end{array}}

\newcommand {\N}{{\cal N}}

\newcommand {\benum} {\begin{enumerate}}
\newcommand {\eenum} {\end{enumerate}}

\newcommand {\bdesc} {\begin{description}}
\newcommand {\edesc} {\end{description}}

\newcommand {\bfig}[2] {\begin{figure}[htbp]
                        \centerline {
                         \epsfig{figure={#1},clip=,width={#2}}}}
\newcommand {\brotatefig}[2] {\begin{figure}[htbp]
                        \centerline {
                         \epsfig{figure={#1},clip=,angle=-90,width={#2}}}}

\newcommand {\bfigfirst}[2] {\begin{figure}[h]
                        \centerline {
                        \setlength{\epsfxsize}{#2}
                        \epsffile{#1}}}
\newcommand {\efig}[2]{ \caption{#2}
                        \label{fig:#1}
                        \end{figure}
                        \mymarginpar{fig:#1}}
\newcommand {\erotatefig}[2]{ \caption{#2}
                        \label{fig:#1}
                        \end{figure}
                        \mymarginpar{fig:#1}}

\newcommand {\btab}[1]{
                       \begin{table}
                       \centering
                       \begin{tabular}{#1}}
\newcommand {\etab}[3] {
                       \end{tabular}
                       \caption[#3]{#2}
                       \label{tab:#1}
                       \end{table}
                       \mymarginpar{tab:#1}
                       \vspace{.1in}}

\newcommand {\btabular}[1]{\begin{center}
                       \begin{tabular}{#1}}
\newcommand {\etabular}{\end{tabular}
                       \end{center}}

\newcommand {\bdefin}[1]{\begin{definition}
                      \label{def:#1} }
\newcommand {\edefin}       {\end{definition}}

\newcommand {\bassum}[1]{\begin{assumption}
                      \label{ass:#1} }
\newcommand {\eassum}       {\end{assumption}}

\newcommand {\bpro}[1]{\begin{property}
                      \label{pro:#1} }
\newcommand {\epro}   {\end{property}}

\newcommand {\bprop}[1]{\begin{proposition}
                      \label{prop:#1} }
\newcommand {\eprop}       {\end{proposition}}
\newcommand {\rprop}[1]{Proposition \ref{prop:#1}}

\newcommand {\blem}[1]{\begin{lemma}
                      \label{lem:#1} }
\newcommand {\elem}   {\end{lemma}}

\newcommand {\bthe}[1]{\begin{theorem}
                      \label{the:#1} }
\newcommand {\ethe}   {\end{theorem}}

\newcommand {\bcor}[1]{\begin{corollary}
                      \mymarginpar{cor:#1}
                      \label{cor:#1} }
\newcommand {\ecor}   {\end{corollary}}

\newcommand {\bax}[1]{\begin{axiom}
                      \mymarginpar{ax:#1}
                      \label{ax:#1} }
\newcommand {\eax}       {\vspace{-.1in} \end{axiom}}

\newcommand {\bconj}[1]{\begin{conjecture}
                      \mymarginpar{conj:#1}
                      \label{conj:#1} }
\newcommand {\econj}       {\end{conjecture}}

\newcommand {\bex}[2]{\vspace{.1in}
                      \begin{example}
                      \mymarginpar{ex:#1}
                       {\bf #2}
                      \label{ex:#1} \em}
\newcommand {\eex}       {\end{example} \vspace{.3cm} }

\newcommand {\brem}[1]{\begin{remark}
                      \mymarginpar{rem:#1}
                      \label{rem:#1} \em }
\newcommand {\erem}   {\end{remark}}

\newcommand{\beq}[1]{%
\begin{equation}\label{eq:#1}%
}
\newcommand {\beqno}[1]{
                      \begin{eqnarray}
                      \nonumber}

\newcommand {\eeq}       {\end{equation}}
\newcommand {\eeqno}       { && \end{eqnarray}}
\newcommand {\req}[1]{(\ref{eq:#1})}

\newcommand {\bear}[1]{\mymarginpar{eq:#1}
                       \begin{eqnarray}
                       \label{eq:#1} }

\newcommand {\bearno}[1]{\mymarginpar{eq:#1}
                       \begin{eqnarray}
                       \nonumber}

\newcommand {\eear}{\end{eqnarray}}
\newcommand {\eearno}{\end{eqnarray}}
\newcommand {\bsel}{\left \{ \begin{array}{cl}}
\newcommand {\esel}{\end{array} \right.}

\newcommand {\bmat}[1]{\left [ \begin{array}{#1}}
\newcommand {\emat}{\end{array} \right ]}

\def\R{I\kern-0.30em R}
\def\N{I\kern-0.30em N}
\def\P{I\kern-0.30em P}

\newcommand {\bxfig}[2] {\begin{figure}[htbp]
                        \centerline {
                         \includegraphics[width=#2]{#1}}}
\newcommand {\brotatexfig}[2] {\begin{figure}[htbp]
                        \centerline {
                         \includegraphics[width=#2,angle=90]{#1}}}
\DeclareGraphicsExtensions{.pdf,.jpg,.png}

\def\bfa{{\mbox{\boldmath $a$}}}
\def\bfsa{{\mbox{\boldmath\scriptsize $a$}}}

\def\bfe{{\mbox{\boldmath $e$}}}

\def\bfg{{\mbox{\boldmath $g$}}}

\def\bfsg{{\mbox{\boldmath\scriptsize $g$}}}

\def\bfo{{\mbox{\boldmath $o$}}}

\def\bfQ{{\mbox{\boldmath $Q$}}}

\def\bfs{{\mbox{\boldmath $s$}}}
\def\bft{{\mbox{\boldmath $t$}}}
\def\bfu{{\mbox{\boldmath $u$}}}
\def\bfv{{\mbox{\boldmath $v$}}}

\def\bfx{{\mbox{\boldmath $x$}}}
\def\bfy{{\mbox{\boldmath $y$}}}

\def\bfQ{{\mbox{\boldmath $Q$}}}

\def\bfpi{{\mbox{\boldmath $\pi$}}}

\def\calA{{\cal A}}
\def\calG{{\cal G}}

\def\calN{{\cal N}}
\def\calO{{\cal O}}

\def\calS{{\cal S}}

\def\ie{{\em i.e.}\ }

\usepackage{epsfig, epsf, array, latexsym, graphics, multirow, color}
\usepackage{booktabs}
\usepackage{multirow}
\usepackage{multicol}
\usepackage{colortbl}
\usepackage{graphicx}
\usepackage{amsmath}
\usepackage{kpfonts}
\usepackage{algorithm}
\usepackage{algpseudocode}

\def\ie{\textit{i.e.}}

\def\BibTeX{{\rm B\kern-.05em{\sc i\kern-.025em b}\kern-.08em
    T\kern-.1667em\lower.7ex\hbox{E}\kern-.125emX}}
\begin{document}

\title{DelAC: A Multi-agent Reinforcement Learning of Team-Symmetric Stochastic Games}
\author{Duan-Shin Lee~\IEEEmembership{Senior Member,~IEEE}, 
	Yu-Hsiu Hung,  
	\IEEEcompsocitemizethanks{\IEEEcompsocthanksitem 
		D.-S. Lee 
		is with the Department of Computer Science, 
		National Tsing Hua University.
		(Email: lds@cs.nthu.edu.tw). Yu-Shin Hung is with MediaTek Inc. 
		(Email: forcer.hung@mediatek.com)}
	\thanks{This research was supported in part by the Ministry of Science and Technology,
		Taiwan, R.O.C., under Contract 114-2221-E-007-080.}} 
\IEEEtitleabstractindextext{
\begin{abstract}
In this paper we study team-symmetric games with $m\ge 2$ teams.  
Players within a team have symmetric identity and have a common payoff 
function.  We show that team-symmetric games always have a team-symmetric 
Nash equilibrium.  We develop and solve a linear complementarity problem 
of team-symmetric Nash equilibria.  We propose an actor-critic based 
multi-agent reinforcement learning algorithm for team-symmetric games. 
Through simulations, we show that this multi-agent reinforcement learning 
algorithm performs much better than many existing algorithms.	
\end{abstract}
}

\maketitle

\IEEEdisplaynontitleabstractindextext
\IEEEpeerreviewmaketitle

\section{Introduction}
\label{sec:Introduction}
For a long time games have been a versatile tool to model various types of interactions among multiple entities including collaboration and competition. Multi-agent reinforcement learning (MARL) emerges as a powerful tool to train agents to learn their equilibrium strategies. There is a rich literature on MARL \cite{Albrecht2024}. In an MARL model, interactions among agents are typically governed by a stochastic game.  One of the most widely accepted solution concepts of games is the Nash equilibrium.  Indeed, many early research results on MARL have focused on training agents to play according to Nash equilibria of the game as a goal \cite{Shoham2007} \cite{Hu2003}. However, recently \cite{Daskalakis2006} and \cite{Chen2006} showed that computing Nash equilibria is PPAD-complete. Thus, it is important to develop MARL methods for special games that demand less computational complexity, while important in practice.

In this paper we study the MARL algorithms of team symmetric games. Symmetric games and team games are special games that find many real-world applications, such as cooperative robots, autonomous vehicles, and etc.  We refer the reader to \cite{djehiche2017mean} for applications in engineering, and \cite{sanchez2024efficiency} for an application in epidemic models. Symmetric games refer to games in which players have indistinguishable identities \cite{Nash1951}\cite{Emmons2021}.  Team games usually refer to games in which players have common payoff functions \cite{Emmons2021}. Symmetric games have been widely applied to study problems in economics, biology, social sciences, and engineering. We refer the reader to \cite{lee2025symmetric} and \cite{hefti2017equilibria} for details.  A closely related game is the anonymous games \cite{Daskalakis2007} \cite{Kalai2005}. Theoretic results on symmetric games were reported in \cite{hefti2017equilibria} and \cite{lee2025symmetric}. Hefti in \cite{hefti2017equilibria} studied symmetric equilibria and asymmetric equilibria of a symmetric games and pointed out a condition for asymmetric equilibria to exist. 

In many real-world multi-agent systems, agents are not merely independent entities 
but are naturally clustered into functional groups with identical characteristics 
and shared objectives. For instance, in swarm robotics for automated logistics, a 
fleet of hardware-identical robots might be assigned to a common task, while
another fleet of robots are assigned to a different task \cite{djehiche2017mean}.
In 5G/6G communication networks, network slicing partitions identical devices into 
distinct service teams such as URLLC and eMBB \cite{Abreu2019}. 
Similarly, in smart grid management, 
thousands of identical residential smart meters interact with centralized storage 
systems to optimize energy distribution \cite{Saad2012}. 
In a network security context, interactions 
often occur between groups of identical agents.
An attacking "Team" may consist of identical malware bots attempting to penetrate 
a network. Since the bots in a botnet share a common objective and are often clones 
of the same code, they are perfectly symmetric agents.  Defending "Team" consists 
of identical automated firewall agents \cite{Do2017}.  
Recent advances in disaggregated LLM inference \cite{arXiv:2602.14516} and 
agentic resource management \cite{arXiv:2603.13110} highlight a transition toward 
self-governing data centers. In these systems, clusters of GPUs with 
indistinguishable identities form specialized teams — such as prefill 
and decode clusters — that must 
autonomously navigate trade-offs between computational throughput and 
memory bandwidth \cite{arXiv:2601.22001}.  In these high-dimensional environments, 
general-purpose MARL algorithms often fail to scale due to the exponential growth 
of joint action spaces and the inherent non-stationarity of 
independent learners \cite{Daskalakis2007}. 
Exploiting the inherent team symmetry of these systems is not merely a simplification, 
but a strategic necessity to achieve computational tractability and stable 
convergence toward equilibrium-consistent strategies.
In this paper we study a symmetric game with $m$ teams.  Specifically, players in 
the same team have indistinguishable identities and have a common payoff function. 
This game can model cooperation within teams and competition between teams. We 
call this game a team-symmetric game.  
One contribution of this paper is to show that team-symmetric games always have 
a team-symmetric Nash equilibrium.  Because of the team symmetry
property, the linear complementarity problem for the 
team-symmetric Nash equilibria has much lower complexity.  Another contribution of
this paper is that we develop an MARL algorithm for a stochastic team-symmetric game.  
Our algorithm is based on actor-critic architecture and numerical solution of 
linear complementarity problem of the game.  Exploiting the team symmetric
property, this MARL algorithm has much lower computational complexity than
any algorithm that does not exploit the property. Through simulations 
we show that our algorithm performs much better than many existing 
MARL algorithms in the literature if the underlying stochastic game 
has the team-symmetric property. 

The outline of this paper is as follows.  In Section~\ref{sec:tsg}
we present team-symmetric games and show that such games always
have symmetric Nash equilibria.  In Section~\ref{sec:computation}
we present a computation procedure for a symmetric Nash equilibrium
of a team-symmetric game. In Section~\ref{sec:DelAC} we present
DelAC algorithm.  We present simulation results in Section~\ref{sec:Experiments}
and conclusions of this paper in Section~\ref{sec:Conclusion}.

\section{Team-Symmetric Games}
\label{sec:tsg}
We consider a stochastic normal-form game $\calG=(\calN,\calO,\calA,u,P,\gamma)$, where $\calN$ 
is a set of players, $\calA$ is a set of actions available 
to the players, $\calO$ is 
a local observations, $u$ is payoffs, $P$ is a transition 
probability function, and 
$\gamma$ is a discount factor in $[0, 1)$.  In this paper 
we assume that there are $m$ teams, where $m\ge 2$.
Team $i$ has $n_i$ players for $i= 1, 2, \ldots m$.  We label the players such that
\[
\calN_i=\left\{k\ \Biggl|\ \sum_{j=1}^{i-1} n_j+1 \le k\le \sum_{j=1}^{i} n_j\right\}
\] 
for agents in team $i$. Thus, $\calN=(\calN_1, \calN_2, \ldots, \calN_m)$.  Let $n$
denote the total number of agents, \ie
\[
n=n_1+n_2+\ldots+n_m.
\]
We assume that players in the same team
have the same action spaces.
Let $\calA_i$ be the action space of a player in team $i$. Thus,
$\calA=(\calA_1)^{n_1}\times\cdots\times (\calA_m)^{n_m}$ is the product of action spaces 
of players in the $m$ teams. Without loss of generality, we assume that
$\calA=\{1, 2, \ldots, |\calA|\}$.
In this paper we use boldface letters to
denote vectors and matrices.  Let
$\bfu=(u_1, u_2, \ldots, u_{n})$ be payoff functions, 
where $u_i: \calA \rightarrow R$ is the
payoff for each player $i$.  Let $\Gamma(\bfu)$ denote a normal-form game with
payoff functions $\bfu$.  We present a few definitions.

\bdefin{cpwt}
We say that $\calG$ has common-payoff within teams 
if 
\beq{common-payoff}
u_i(\bfa)=u_j(\bfa), \mbox{for all $i, j\in \calN_k$ and for all $k=1, 2, \ldots, m$}
\eeq 
and for all action profiles $\bfa\in \calA$. 
\edefin

\bdefin{pwt}
A permutation
within teams $\phi$ is a bijection from $\calN$ to itself 
with additional properties that
$\phi(i)\in \calN_k$ for any $i\in \calN_k$ and 
for any $1\le k\le m$.  
\edefin

\bdefin{stg}
Suppose that player $i$ takes action $a_i$.
We call $\calG$ a team-symmetric game if 
\beq{symmetric-game}
u_{\phi(i)}(a_1, a_2, \ldots, a_{n})
= u_i(a_{\phi(1)}, a_{\phi(2)}, \ldots, a_{\phi(n)})
\eeq
for all $(a_1, a_2, \ldots, a_{n})\in\calA$, $1\le i\le n$.
\edefin

Note that \req{common-payoff} and \req{symmetric-game} imply that
\beq{cps}
u_{i}(a_1, a_2, \ldots, a_{|\calN|})
=u_{\phi(i)}(a_{\phi(1)},a_{\phi(2)}, \ldots, a_{\phi(n)}).
\eeq

We represent mixed strategies by $|\calA|$ dimensional real vectors.  Specifically,
let $\bfv_{ia}$ denote an $|\calA|$ dimensional vector, whose entries are all
zero, except that the $a$-th entry is one.  $\bfv_{ia}$ denotes the pure strategy
$a$ of player $i$.    
We use notation $\bfs_i$ to denote a mixed strategy of player $i$.  
The $a$-th entry of $\bfs_i$ is the probability that player $i$ uses action 
$a$.  Mixed strategy $\bfs_i$ is a point in a simplex with vertices $\{\bfv_{ia}\}$.
We denote the expected payoff of player $i$
when strategy profile $\bfs=(\bfs_1, \bfs_2, \ldots, \bfs_{n})$ is used by
\beq{exppay}
\bar u_i(\bfs_1, \bfs_2, \ldots, \bfs_{n})=
\sum_{(\bfs_1, \bfs_2, \ldots, \bfs_{n})} u_i(\bfa)\prod_{j=1}^{n} (\bfs_j)_{a_j},
\eeq
where $(\bfs_j)_{a_j}$ denotes the $(a_j)$-th entry of $\bfs_j$ and is the 
probability that action $a_j$ is used by player $j$.  
We overload notations and use $u_i$ to also denote the expected payoff of player $i$. 
We adopt a standard notation that $\bfs_{-i}=(s_1, \ldots, s_{i-1}, s_{i+1}, \ldots, s_{n})$.
Let $(\bfs_{-i}, \bfs^\prime)$ denote
\[
(\bfs_1, \ldots, \bfs_{i-1}, \bfs^\prime, \bfs_{i+1}, \ldots, \bfs_{n}),
\]
\ie\ player $i$'s strategy in profile $\bfs$ is replaced by $\bfs^\prime$ 
and other players' strategies are unchanged.  
Note that \req{cps} can be extended to mixed strategies, \ie\ 
\beq{tss}
\begin{aligned}
&u_{i}(\bfs_1, \bfs_2, \ldots, \bfs_{n})\\
&=u_{\phi(i)}(
\bfs_{\phi(i)},\bfs_{\phi(2)}, \ldots, \bfs_{\phi(n)})
\end{aligned}
\eeq
Also note that $\phi$ is a permutation within teams on players.  It can induce
a permutation on strategies.  Specifically, define permutation $\rho$ such 
that
\[
\rho(\bfs)=(\bfs_{\phi(i)},\bfs_{\phi(2)}, \ldots, \bfs_{\phi(n)}).
\]
\bdefin{sst}
We call strategy profile $\bfs$ team symmetric or symmetric within teams if
\beq{ssp}
\rho(\bfs)=\bfs
\eeq
for all permutation $\rho$ within teams.
\edefin
One main result of this paper is the following proposition. 
Its proof is presented in Appendix.
\bprop{sne}
A team-symmetric game with common payoffs has a team-symmetric Nash equilibrium.
\eprop
\section{Computation of team-symmetric Nash equilibria}
\label{sec:computation}
In this section, we formulate a feasibility program to numerically compute a team symmetric
Nash equilibrium of a game with $m$ teams and common payoffs within teams.  We consider
strategy profile of the form
\begin{equation*}
	\bfs=(\underbrace{x_1, \ldots, x_1}_{n_1}, 
	\underbrace{x_2, \ldots, x_2}_{n_2}, 
	\ldots,
	\underbrace{x_m, \ldots, x_m}_{n_m}),
\end{equation*}
where $(\bfx_i, \ldots, \bfx_i)$ is the strategy profile used by members in team $i$ for
$1\le i\le m$. Since strategy profiles are symmetric within teams, 
we assume that $\bfx_i$ is a mixed strategy adopted by players 
in team $i$.  Particularly, the $j$-th entry of vector $\bfx_i$, denoted by $(\bfx_i)_j$, 
is the probability that an agent in team $i$ plays action $j$. We let
\[
u_i(\bfx_1, \bfx_2, \ldots, \bfx_m)
\]
denote the payoff of an agent in team $i$. 
We overload the notation and let
\[
u_{i,j}(\bfx_1, \bfx_2, \ldots, \bfx_m)
\]
be the payoff of an agent in team $i$ who plays action $j$ and all agents in
team $k$ play mixed strategy $\bfx_k$ for $k\ne i$ and all other agents
in team $i$ play $\bfx_i$. We also overload the notation
\[
\bar u_{i,j}(\bfx_1, \bfx_2, \ldots, \bfx_m)
\]
to denote the expected value of $u_{i,j}(\bfx_1, \bfx_2, \ldots, \bfx_m)$.

Since we assume that game $\calG$ is team
symmetric and has common payoffs, for each realization of $\bfx_i$, where $1\le i\le m$,
we let $\bfg_i$ denote the number of agents who play an action in team $i$.  Specifically,
let $(\bfg_i)_j$ be the number of agents who play action $j$ for $1\le j\le |\calA|$
in team $i$. For each realization of $(\bfx_1, \bfx_2, \ldots, \bfx_m)$, 
there is a corresponding vector $(\bfg_1, \bfg_2, \ldots, \bfg_m)$. 
We overload the notation again and let
\[
u_i(\bfg_1, \ldots, \bfg_{i-1}, \bfg_i+\bfe_a, 
\bfg_{i+1},\ldots, \bfg_m)
\]
be the payoff of an agent in team $i$ who plays action $a$, where all other agents
in team $j$ plays mixed strategy $\bfg_j$, $1\le j\le m$.
Since players sample actions independently according to their mixed strategies,
vector $\bfg_j$ occurs according to a multinomial distribution
with parameters $n_j$ and $\bfx_j$ for $j\ne i$. For team $i$, vector
$\bfg_i$ occurs also according to a multinomial distribution.  However, since
one agent plays action $a$, the parameters in this case are $n_i-1$ and $\bfx_i$.
The expected payoff of team $i$ as a function of team strategies $\bfx$
and $\bfy$ is shown in \eqref{long-eq}.
\begin{figure*}[t] 
	\begin{align}
		\bar u_{i,a}(\bfx_1,\ldots, \bfx_m)&=
		\sum_{\{\bfsg_1 | \sum (\bfsg_1)_\ell=n_1\}}\ldots
		\sum_{\{\bfsg_i | \sum (\bfsg_i)_\ell=n_i-1\}}\ldots 
		\sum_{\{\bfsg_m|\sum (\bfsg_m)_\ell=n_m\}} 
		\frac{(n_i-1)!
		\prod_{k=1}^{|\calA|} (\bfx_i)^{(\bfg_i)_k}}{\prod_{k=1}^{n_i}
		(\bfg_i)_k!} 
		\prod_{\substack{j\ne i \\ j=1}}^{m}\frac{n_j! \prod_{k=1}^{|\calA|} 
			(\bfx_j)^{(\bfg_j)_k}}{\prod_{k=1}^{n_i}(\bfg_i)_k!}\nonumber\\
		&\qquad	\cdot u_i(\bfg_1, \ldots, \bfg_{i-1}, \bfg_i+\bfe_a, 
			\bfg_{i+1},\ldots, \bfg_m).\label{long-eq}
	\end{align}
\end{figure*}
In \eqref{long-eq} $\bfe_a$ is vector whose entries are all zero, 
except that the $a$-th entry is one.
We consider a special case, in which payoffs are linear in the number of
agents who play actions.  Specifically,
\[
u_i(\bfg_1, \bfg_2, \ldots, \bfg_m)=\sum_{k=1}^m \sum_{j=1}^{|\calA|} c_{k,j}(\bfg_k)_j
\]
for some constants $\{c_{k,j}: 1\le k\le m, 1\le j\le |\calA|\}$.
In this special case, $\tilde u_{i,a}$
can be further simplified as
\beq{simplified-payoff}
\tilde u_{i,a}(\bfx,\bfy)=c_{i,a}+\sum_{j=1}^{|\calA|}(n_i-1)c_{i,j}(\bfx_i)_j+
\sum_{k=1, k\ne i}^m \sum_{j=1}^{|\calA|}n_k c_{k,j}(\bfx_k)_j.
\eeq

It is well known that Nash equilibria can be numerically obtained by solving 
a linear complementarity problem (LCP) \cite{Shoham2009}.  This LCP can be viewed
as a nonlinear program without objective functions, \ie a feasibility program.
The feasibility program for the
team-symmetric games with common payoffs is significantly simpler than that of 
a typical general-sum game.  We present the program for the
sake of completeness.  
\begin{align}\label{fp}
	&\tilde u_{i,j}(\bfx_1,\ldots,\bfx_m)+r_{i,j} \le U_{i,j}^*,\  
	1\le i\le m, 1\le j\le |\calA|, \\
	&\sum_{j=1}^{|\calA|} (\bfx_i)_j=1, \ 1\le i\le m, \nonumber\\
	&(\bfx_i)_j\ge 0, r_{i,j}\ge 0,\ (\bfx_i)_j\cdot r_{i,j}=0,\ 1\le i\le m, 1\le j\le
	 |\calA|.\nonumber
\end{align}

The computational advantage of the team-symmetric approach is most evident 
when comparing its complexity to general $n$-player games. In a 
standard $n$-player MARL framework, the feasibility program required 
to find a Nash equilibrium involves $n(2|\mathcal{A}|+1)$ variables, 
with expected payoff calculations requiring a summation over an exponential 
joint action space of size $|\mathcal{A}|^n$ . In contrast, our team-symmetric 
formulation in \eqref{fp} reduces the variable count to $m(2|\mathcal{A}|+1)$, 
where $m \ll n$. Furthermore, Eq. \eqref{long-eq} replaces the exponential 
joint-action summation with a combination-based sum over team-wise action 
counts, shifting the complexity from exponential in $n$ to polynomial in 
$n$. This transition from exponential 
to group-based scaling is a strategic necessity for achieving tractability 
in large-scale systems like swarm robotics and data center management..

\section{Delegate Actor-Critic Networks}
\label{sec:DelAC}
In this section we present our reinforcement learning algorithm to compute a
symmetric Nash equilibrium of a team-symmetric stochastic game.  
Actor-critic algorithms form a paradigm for multiagent reinforcement learning of
stochastic games.  In this framework, an actor is a neural network that trains a
parameterized policy based on policy gradient algorithms.  In addition, a neural
network called critic is introduced in this frame to estimate the state-value of the
game.  We further remark that actor-critic methods follow a centralized training
and distributed execution (CTDE) paradigm.

We first describe a baseline algorithm.  We then simplify the algorithm.  The
simplified version of the algorithm will be called delegate actor-critic (DelAC) algorithm.
We allocate one actor and one critic to each player.  Initially, all actors generate
the same mixed strategy distributions, and all critic networks generate zero $Q$ values.
At time $t$, the actor 
associated with player $i$ generates a mixed strategy for player $i$, where
$1\le i\le n$.  Player $i$ samples an action $a_i^t$ from this mixed strategy.
An action profile $\bfa^t=(a_1^t, a_2^t, \ldots, a_{n}^t)$ is executed,
and the environment generates observations $\bfo^{t+1}$ and rewards $\bfu^t$.
The discounted reward of player $i$ from time
$t$ onward is defined as
\beq{dr}
d_i^t= \sum_{\tau=t}^\infty \gamma^{\tau-t} u_i^{\tau},
\eeq
where $\gamma$ is a discount factor in interval $(0, 1)$, and $u_i^t$ is the reward
received by player $i$ at time $t$.  We remark that the superscript of $\gamma^\tau$ in 
\req{dr} denotes powers, and the superscripts in $d_i^t$ and $u_i^t$ indicate time.
We also remark that we overload the symbol $u$ here. 
The action-value function $Q_i(s,a)$ for player $i$,
\beq{avf}
Q_i^\pi(s, a)=E\left[d_i^t | s^t=s, a_i^t=a\right],
\eeq
is the expected discounted reward when selecting action $a$ in state $s$ and 
following policy $\pi$. 

Critic networks are trained to generate action-value functions using 
$\{(\bfo^\tau, \bfo^{\tau+1}, \bfa^\tau, \bfu^\tau), 1\le\tau \le t\}$. 
Overloading symbol $Q$, we denote the output of critic $i$ by $Q_i(\bfo^t, \bfa^t)$
in order to emphasize that it is trained with $\{(\bfo^\tau, \bfa^\tau), 1\le\tau \le t\}$.
We also emphasize that all critics are trained with the same data and in the
same order.  We construct a normal-form game $\Gamma(\bfQ(\bfo^t, \bfa^t))$,
in which payoffs of player $i$ is $Q_i(\bfo^t, \bfa^t)$ for $1\le i\le n$. 
According to \rprop{conj1} to be presented below, game $\Gamma(\bfQ(\bfo^t, \bfa^t))$ is a
team-symmetric game.
We solve a Nash equilibrium of the game using (\ref{fp}).
Denote the solution of (\ref{fp}) by $\{\tilde\pi_i^{t}, 1\le i\le n\}$.
The actor network of player $i$ attempts to learn $\tilde\pi_i^{t}$ by minimizing loss 
function
\beq{actor-loss}
\mathcal{L}_i(t)=\sum_{\tau=1}^t D_{\text{KL}} \left( \tilde\pi_i^{\tau} 
\parallel \pi_i(\cdot \mid o_i^t) \right),
\eeq
where $D_{\text{KL}}(\cdot)$ denotes the KL divergence.

The actor associated with
player $i$ attempts to learn player $i$'s mixed strategy in a symmetric Nash
equilibrium within teams.  
The $i$-th critic attempts to learn the
action-value function $Q_i^\pi(s, a)$ of player $i$.  The $i$-th critic 
learns $Q_i(s,a)$ by minimizing loss function
\[
L(t)=\sum_{\tau=1}^t\sum_{i=1}^{n} \left(Q_i(\bfo^\tau, \bfa^\tau)-y_i^\tau\right)^2,
\]
where $y_i^t$ is a sample value of $d_i^t$ in \req{dr}.  

The proof of \rprop{conj1} is quite simple and is presented in the Appendix.
\bprop{conj1}
\begin{enumerate}
	\item Actor networks produce symmetric mixed strategy distributions within teams,
	\ie\ $\pi_j(\cdot | o_j)=\phi_k(\cdot | o_k)$ for $j, k\in \calN_i$, $j\ne k$
	and for any $1\le i\le m$.
	\item The normal-form game $\Gamma(\bfQ(\bfo^t, \bfa^t))$
	produced by the critic is symmetric within teams, \ie\  
	$Q_j(\bfo^t,\bfa^t)=Q_k(\bfo^t,\bfa^t)$ for $j, k\in \calN_i$, $j\ne k$
	and for any $1\le i\le m$.
\end{enumerate}
\eprop

Due to \rprop{conj1}, each team needs only one delegate actor and one delegate
critic.  In fact, the $m$ delegate critic networks can be merged into one network.
We present the DelAC algorithm in Algorithm \ref{algo:DelAC}.  
In Algorithm \ref{algo:DelAC} we choose actors $1, n_1+1, n_1+n_2+1, 
\ldots, \sum_{i=1}^{m-1}n_i+1$ as the delegates of the $m$ teams.
In the presentation of Algorithm \ref{algo:DelAC}, we overload the symbol $\pi_i(\cdot
| o_i)$ to denote the delegate actor network for team $i$.

\begin{algorithm}
\caption{Delegate Actor-Critic Network}
\label{algo:DelAC}
\begin{algorithmic}[1]
\State Initialize $m$ delegate actor networks which 
generate $\pi_1(\cdot | \bfo^t, \bfu^t)$,
$\pi_{n_1+1}(\cdot | \bfo^t, \bfu^t), \ldots, 
\pi_{\sum_{j=1}^{m-1} n_j+1}(\cdot | \bfo^t, \bfu^t)$
\State Initialize centralized critic which generates $Q(\bfo^t, \bfu^t)$
\For{each episode from $1$ to $M$}
    \State Obtain initial observations $(o_1, \dots, o_N)$
    \For{each time step $t = 1$ to $T$}
        \For{each agent $i = 1$ to $n$}
            \State Sample action $a_i^t \sim \pi_j(\cdot | \bfo^t,\bfu^t)$ for $i\in \calN_j$,
            \Statex \hspace{2cm}  $1\le j\le m$
        \EndFor
        \State Form joint action $\bfa^t = (a_1^t, \dots, a_{n}^t)$
        \State Execute $\bfa^t$
        \State Receive payoffs $\bfu^t$ and observations $\bfo^{t+1}$ of 
        \Statex \hspace{1.5cm} the next time step
        \State Store $(\bfo^t, \bfu^t, \bfa^t, \bfo^{t+1})$ into a batch
    \EndFor
    \State Convert each joint action profile $\bfa^t$ in a batch 
    \Statex \hspace{1cm} into team-wise action count vector 
    \Statex \hspace{1cm} $(\bfg_1^t, \bfg_2^t, \ldots, \bfg_m^t)$
    \State Compute targets for each time step in a batch:
    \beq{critic-target}
        y_j^t = \sum_{k=t}^T \gamma^{k - t} u_j^k, \quad \text{for } 1\le j\le m
    \eeq
    \State Update critic by minimizing total loss collected 
    \Statex \hspace{1cm} in a batch:
    \[
        L(\theta) = \sum_{t} \sum_{j=1}^m \left(Q_j(\bfo^t,\bfa^t) - y_j^t \right)^2
    \]
    \For{each $t$ in batch}
        \State Construct game $\Gamma(\bfQ({\bfo^t},\bfa^t))$ from the critic
        \State Solve (\ref{fp}) for equilibrium $\tilde\bfpi^{t}$ from $\Gamma(\bfQ({\bfo^t},\bfa^t))$ 
        \For{each agent $i = 1$ to $n$}
            \State Accumulate KL divergence loss in \req{actor-loss}
        \EndFor
    \EndFor
    \For{each agent $i = 1$ to $n$}
        \State Update actor $\phi_i$ using gradient descent on $\mathcal{L}_i$
    \EndFor
\EndFor
\end{algorithmic}
\end{algorithm}

\section{Experiments}
\label{sec:Experiments}
In this section we present simulation of Algorithm 1 on several team-symmetric games.
We compare the performance of Algorithm \ref{algo:DelAC} with those of 
several well known MARL algorithms in the literature. We present a brief 
review of these baseline algorithms in Section~\ref{sec:baselines}. 
The performance of Algorithm \ref{algo:DelAC} is presented in Section~\ref{sec:mr}.


\subsection{Review of Some MADRL Algorithms}
\label{sec:baselines}
In this section we review a few well known MARL algorithms, with which the performance
of Algorithm \ref{algo:DelAC} will be compared.  These algorithms can be broadly
classified into two categories, value based and actor-critic based methods.
All baseline methods considered are compatible with the centralized training 
with decentralized execution (CTDE) paradigm, unless otherwise noted.

Independent Q-learning (IQL) is a value based algorithm which treats agents as
independent learners \cite{mnih2013playing}. Each agent independently determines
his/her action without coordinating with other agents.  Nash Q-learning (NashQ) 
\cite{hu2003nash} is another value based method.  Nash Q-learning 
approximates payoffs of a stochastic
game by Q-values and computes a corresponding Nash equilibrium.  
Friend-or-Foe Q-learning (FFQ) classifies agents into a set of friends and a set of foes.
FFQ applies a min-max operation to all foes and a max-sum operation to friends
\cite{littman2001friend}.   QMIX~\cite{rashid2018qmix} is a value decomposition
method developed for cooperative settings.   It factorizes a centralized action-value function 
into individual per-agent Q-functions using a monotonic mixing network.
NWQMIX~\cite{CY-Wu2024} extends QMIX to environments where both cooperation and
competition exist.  For cooperative agents, original factorization proposed by QMIX
is retained.  For competing agents, new factorization is proposed.  

For actor-critic based methods, we compare Algorithm \ref{algo:DelAC} with
Independent Advantage Actor-Critic (IA2C)~\cite{lowe2017multi}, 
{Independent Proximal Policy Optimization (IPPO)}~\cite{schulman2017proximal}, \cite{papoudakis2020benchmarking},
{Centralized Advantage Actor-Critic (CA2C)}~\cite{lyu2023centralized}, and
{Multi-Agent Proximal Policy Optimization (MAPPO)}~\cite{yu2022surprising}.
IA2C allocates an actor and a critic to each agent.  Actors and critics
of all agents learn independently without centralized coordination.  
As IA2C, IPPO allocates an actor and a critic to all agents, and these
actors and critics perform proximal optimization independently.
CA2C incorporates a centralized critic that has access to global state and joint action 
information during training.  See Chapter 9.4.3 in \cite{marl-book}.
Both CA2C and MAPPO follow the CTDE paradigm.

We remark that in the implementation of the MARL methods above, we adopt a
parameter sharing architecture \cite{terry2020revisiting} and 
\cite{christianos2021scaling}. Agents within the same team share the same 
value network and policy network. Despite sharing parameters, agents retain independent 
observations and actions, allowing for decentralized execution during both training and evaluation.

\subsection{Results}
\label{sec:mr}
Although our work holds for general $m$, in numerical studies we assume
that $m=2$. We simulate 30 randomly generated two-team symmetric games.
In these games, each team has two players and
each player has two actions.  The payoffs are randomly selected integers
in the range $[0, 10]$.  The hyper-parameters used in the experiments are shown
in Table \ref{tab:hp}. 
\begin{table}[ht]
	\centering
	\begin{tabular}{|c|c|c|}\hline
	    Hyper-parameters       & Symbol     & Value      \\ \hline\hline
		Number of training steps     & $T_{total}$    & $50,000$    \\ \hline
		Discount factor  & $\gamma$ & $0.99$    \\ \hline
		GAE Lambda     & $\lambda$    & $0.95$    \\ \hline
		Learning rate & $\alpha$ & $3\times 10^{-4}$ for actor \\
		 & & $3\times 10^{-2}$ for critic \\ \hline
		PPO clip parameter & $\epsilon_{clip}$ & 0.2 \\ \hline
		Optimization epochs & $K$ & 4 \\ \hline
		Batch size & $B$ & 256 \\ \hline
		Entropy coefficient &  & 0 \\ \hline
		Max gradient norm &  & 0.5 \\ \hline
		Parallel environments & $N_{env}$ & 4 \\ \hline
	\end{tabular}
	\caption{Hyper-parameters of the experiments.}
	\label{tab:hp}
\end{table}

\begin{figure}[htbp]
	\centering
	\begin{subfigure}[b]{\linewidth}
	\centering
	\includegraphics[width=1\textwidth]{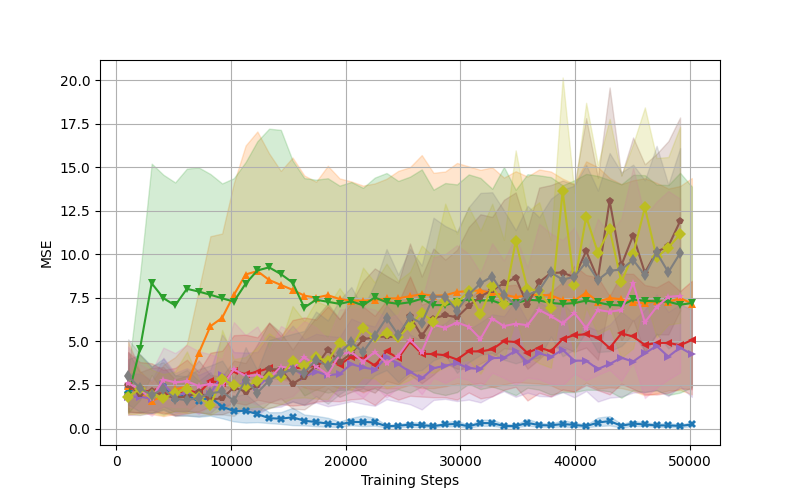}
	\caption{\ \vspace{0.5cm}}
\end{subfigure}
\vspace{1em}  
\begin{subfigure}[b]{\linewidth}
	\centering
	\includegraphics[width=1\textwidth]{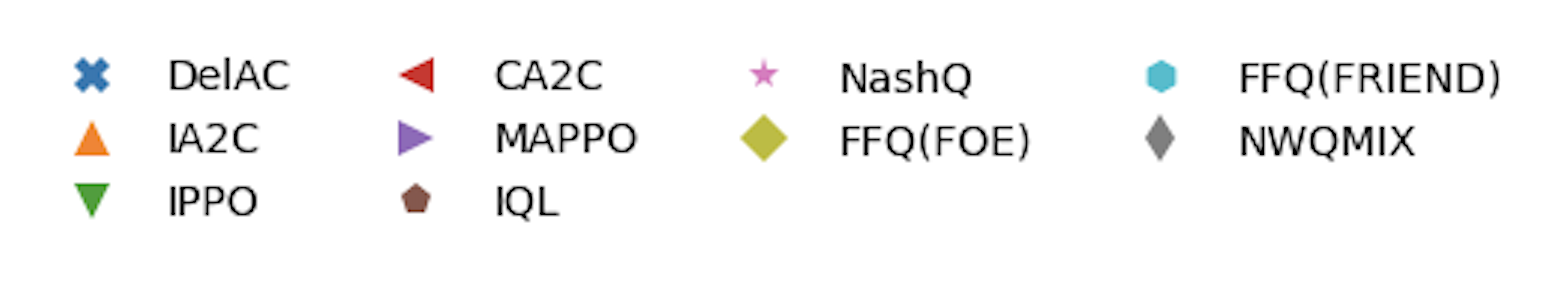}
	\caption{\ }
\end{subfigure}
\caption{Panel (a) contains the average MSE of 30 random symmetric zero-sum games with two teams. 
Panel (b) contains curve legends of the curves in panel (a).}\label{result_zerosum}
\end{figure}

\begin{figure}[htbp]
	\centering
	\includegraphics[width=1\textwidth]{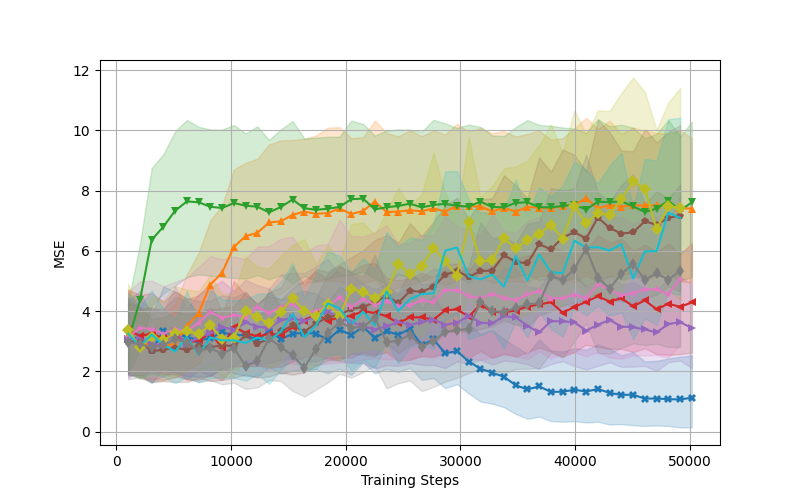}
	\caption{Average MSE of 30 random symmetric, general-sum games with two teams.}
	\label{result_generalsum}
\end{figure}

We compare the average mean squared error (MSE) 
of DelAC and the MARL algorithms described in Section \ref{sec:baselines}. 
In each time step, we compute the average of the two team MSEs. A team MSE is
defined as an MSE between agents' expected payoffs and the expected payoffs 
corresponding to a symmetric Nash equilibrium.  Fig.~\ref{result_zerosum}
contains the average MSE over 30 random team-symmetric zero-sum games.
Panel (b) of Fig.~\ref{result_zerosum} shows the curve legends of MARL methods.
The same legends will be used in subsequent figures.
Fig.~\ref{result_generalsum} contains the average MSE over 30 random team-symmetric
general-sum games.  From these two figures, we see that DelAC has the 
least average MSE compared with other methods.

\begin{figure}[htbp]
	\centering
	\includegraphics[width=1\textwidth]{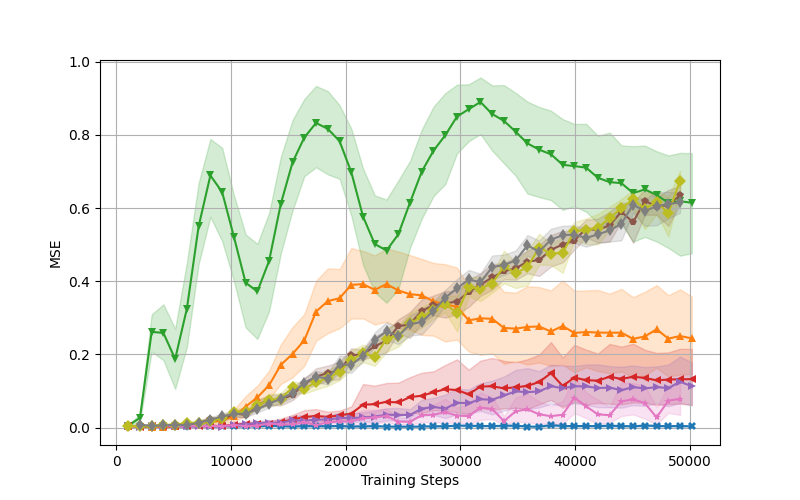}
	\caption{MSE of GMP game with $\omega$=0.5.}
	\label{result_gmp}
\end{figure}

We have also simulated a benchmark game called the generalized matching penny
(GMP) game.  The payoff matrix of this game is shown in Table \ref{tab:gmp}.
In a GMP, there are two teams and each team has two agents.  Each agent
has two actions, \ie\ action H and action T.  From Table \ref{tab:gmp},
a GMP is a zero-sum game.  According to \cite{kalogiannis2023towards}, 
when the game parameter $\omega$ satisfies $0<\omega<1$, a GMP game 
admits a unique Nash equilibrium, which is a mixed strategy. This property 
eliminates the existence of pure strategy equilibria, thereby establishing 
GMP as a particularly challenging environment for evaluating convergence 
behaviors in mixed-strategy settings. Fig.~\ref{result_gmp} shows DelAC has
least MSE compared with those of other methods.

From Figs.~\ref{result_zerosum}, \ref{result_generalsum} and \ref{result_gmp}
we see that DelAC consistently outperforms other methods.  In the training
process, DelAC maintains a continuously decreasing MSE, while other methods
either converge to non-optimal values or exhibit significant swings.   
\begin{table}[ht]
	\centering
	\begin{tabular}{|c|c|c|c|}\hline
		& HH       & HT/TH      & TT       \\
		\hline
		HH     & $1, -1$    & $\omega, -\omega$ & $-1, 1$    \\
		\hline
		HT/TH  & $-\omega, \omega$ & $0, 0$       & $-\omega, \omega$ \\
		\hline
		TT     & $-1, 1$    & $\omega, -\omega$ & $1, -1$    \\ \hline
	\end{tabular}
	\caption{Payoff matrix of a GMP game.}
	\label{tab:gmp}
\end{table}
We study the average KL divergence between learned strategy profiles and Nash equilibria.
These averages are taken with respect to 30 random games and four players in each game.
The results are shown in Table~\ref{result_kldiv}.  
We see that DelAC has much smaller average KL divergences than other methods.
\begin{table}[htbp]
	\centering
	\begin{tabular}{|l|l|l|l|}
		\hline
		& General-sum & Zero-sum & GMP($\omega$=0.5) \\ \hline \hline
		IA2C & 2.276±4.900 & 1.584±4.396 & 1.303±2.197 \\ \hline
		IPPO & 3.343±7.231 & 2.379±6.722 & 2.591±3.563 \\ \hline
		CA2C & 0.858±2.718 & 0.363±1.634 & 1.128±2.907 \\ \hline
		MAPPO & 0.744±2.431 &0.301±1.575 & 1.005±2.789 \\ \hline
		DelAC & \textbf{0.078±0.563} & \textbf{0.001±0.001} & \textbf{0.001±0.004} \\ \hline
	\end{tabular}
	\caption{Average KL divergence between learned policies and Nash equilibria.}
	\label{result_kldiv}
\end{table}

\section{Conclusions}
\label{sec:Conclusion}
In this paper we studied team-symmetric games with $m\ge 2$ teams.  Players within
the same team have common payoff functions.  We showed that team-symmetric games
always have a team-symmetric Nash equilibrium.  We developed and solve a linear
complementarity problem of team-symmetric Nash equilibria.  We proposed an actor-critic
based multi-agent reinforcement learning algorithm for team-symmetric games.
Through simulations, we showed that this multi-agent reinforcement learning algorithm
performs much better than many existing algorithms.

\appendix
\section{Appendix}

In this appendix we present a proof of \rprop{sne} and \rprop{conj1}.  
The proof of \rprop{sne} is 
quite similar to that of Theorem 2 in \cite{Nash1951}, except that
the permutations involved in this paper are permutations within teams.
The proof shall apply Brouwer's fixed point theorem \cite[Chapter II]{bredon2013topology}. 
For completeness, we quote the theorem as follows.
\bthe{Brower}
{\bf (Brouwer's fixed point theorem)} 
Every continuous function from a nonempty convex compact 
subset $K$ of a Euclidean space to $K$ itself has a fixed point.
\ethe

{\bf Proof of \rprop{sne}.}  Define function $f: R^{n}\rightarrow 
R^{n}$.  Let 
\beq{def-f}
f(\bfs)=\hat\bfs,
\eeq 
where
\beq{f}
\hat\bfs_i=\frac{\bfs_i+\sum_j \max\left(0, u_i(\bfs_{-i}, \bfv_{ij})-u_i(\bfs)\right)\bfv_{ij}}
{1+\sum_j \max\left(0, u_i(\bfs_{-i},\bfv_{ij})-u_i(\bfs)\right)}.
\eeq
It has been shown in the proof of Theorem 1 in \cite{Nash1951} that
$f$ has a fixed point and this fixed point is a Nash equilibrium point.
We now present and prove the following two claims.  
\begin{enumerate}
	\item The set of team symmetric strategy profiles is a compact convex 
	subset of the simplex with vertices
	\[
	\left\{(\bfs_1,\ldots,\bfs_{n}: \sum_{j=1}^{|\calA|} (\bfs_i)_j=1, 
	1\le i\le n\right\}.
	\]
	\item Function $f$ maps a team symmetric strategy profile into
	a team symmetric strategy profile.
\end{enumerate}
It follows from the two claims and Brouwer's fixed point theorem that 
$f$ in \req{f} has a fixed point which is symmetric within teams.

It remains to prove the two claims.
For claim 1, suppose that $\bfs$ and $\bft$ are two team-symmetric strategy
profiles.  Let $w$ be a real number in $[0, 1]$.  Then,
for permutation $\phi$ within teams and its induced permutation $\rho$,
\begin{align*}
	\rho(\bfs) &= \bfs \\
	\rho(\bft) &= \bft,
\end{align*}
and
\begin{align}\label{convex-proof}
	&\rho(w\bfs+(1-w)\bft)\nonumber\\
	 &=\rho(w\bfs_{1}+(1-w)\bft_1,\ldots,
	w\bfs_{n}+(1-w)\bft_{n})\nonumber\\
	&=(w\bfs_{\phi(1)}+(1-w)\bft_{\phi(1)},\ldots,
	w\bfs_{\phi(n)}+(1-w)\bft_{\phi(n)})\nonumber\\
&=w \rho(\bfs)+(1-w)\rho(\bft) \nonumber\\
&=w\bfs+(1-w)\bft.
\end{align}
Thus, the set of team symmetric strategy profiles form a convex set.
Let $\calS$ be the set of team symmetric strategy profiles, \ie\ 
$\calS$ contains all $\bfs$ such that $\rho(\bfs)=\bfs$ for any permutation
$\rho$ within teams.  Endow $\calS$ with a metric
\[
d(\bfs,\bft)=\sum_{i=1}^{n} \mbox{TV}(\bfs_i, \bft_i),
\]
where $\mbox{TV}$ is the total variation distance between two probability
mass functions. Let $\calS^c$ be its complement.  Suppose that 
$\bfs\in\calS^c$.  There must be at least two players, say $i$ and $j$, such
that $\bfs_i\ne \bfs_j$.  Define
\[
\mathbf{N}(\bfs)=\left\{\bft: d(\bfs,\bft)< \mbox{TV}(\bfs_i,\bfs_j)\right\}.
\]
It is clear that $\rho(\bft)\ne\bft$ for any $\bft\in \mathbf{N}(\bfs)$.
Thus, $\mathbf{N}(\bfs)\subset \calS^c$.  It follows that $\bfs$ is an interior
point, and thus, $\calS^c$ is open. 

Now we prove claim 2.  Suppose that $\bfs$ is a team symmetric 
strategy profile, \ie \ 
\beq{sym}
\rho(\bfs)=\bfs.
\eeq
Let $\phi$ be a permutation within teams and let $\rho$ be its induced
permutation on strategy profiles.  Suppose that $\phi(i)=k$.
Then, the $i$-th entry of $f(\bfs)$ is given in \req{f}.
Since the $k$-th entry of $\rho(\bfs)$ is $\hat\bfs_i$, 
it is clear that the $k$-th entry of $f(\rho(\bfs))$ is
\[
\left(f(\rho(\bfs))\right)_k = \hat\bfs_i.
\]
Thus,
\beq{interchange}
f(\rho(\bfs))=\rho(f(\bfs)).
\eeq
We have
\beq{final}
f(\rho(\bfs))=f(\bfs)=\hat\bfs=\rho(\hat\bfs),
\eeq  
where the first equality in \req{final} is due to \req{sym}, the second
equality is due to \req{def-f} and the third equality is due to \req{interchange}.
Thus, $\hat\bfs$ is team symmetric.

Next, we prove \rprop{conj1}.  The proof is quite straightforward, and is based on induction.

{\bf Proof of \rprop{conj1}.} Initially at $t=1$ all actors generate
the same mixed strategy distributions, and all critic networks generate zero $Q$ values.
Thus, the two statements of the proposition hold.  Assume that the two statements
hold for some $t$.  Since the back-propagation training of all critic networks in the same team
uses the same data and in the same order, it follows that all critic networks in the same team
are identical at time $t+1$.  Thus, statement two of the proposition is true at time $t+1$.  
Since the normal-form game at time $t$ has symmetric payoffs within teams, it follows
that the game possess a symmetric Nash equilibrium.  Since the actor networks are symmetric
within teams and are trained by a symmetric strategy profile, it follows that at time $t+1$
the actors remain symmetric within teams.  Thus, statement one holds at time $t+1$.  
We thus, complete the induction procedure.

\bibliographystyle{IEEEtran}
\bibliography{reference}

\end{document}